\documentclass[tightenlines,aps,prl,twocolumn,showpacs,showkeywords,reprint]{revtex4}
\usepackage{mathrsfs, mathbbol}
\usepackage{amssymb}
\usepackage{graphicx}
\usepackage{eufrak}
\usepackage{color}
\usepackage[vcentermath]{youngtab}
\usepackage{verbatim}
\usepackage[matrix,frame,arrow]{xypic}
\input{Qcircuit}

\newcommand{\out}[2]{{\ket{#1}\bra{#2}}}
\newcommand{\inner}[2]{{\left\langle \: #1 \left.\right|  #2 \: \right\rangle}}

\newcommand{\lr}[1]{ \left( #1 \right)}

\newcommand{\bea}{\begin{eqnarray}}
\newcommand{\eea}{\end{eqnarray}}

\newcommand{\com}[2]{ \left[ #1,#2 \right]}

\newcommand{\lra}[1]{ \left| #1 \right|}

\newcommand{\lrb}[1]{ \left[ #1 \right]}

\newcommand{\sumlim}[2]{\sum\limits_{#1}^{#2}}

\def \r{\rho}
\def \k{\kappa}
\def \a{\alpha}

\def \l{\lambda}

\def \s{\sigma}
\def \om{\omega}
\def \O{\Omega}

\def \ah{\hat{a}}
\def \ahd{\hat{a}^{\dagger}}

\def \da{\downarrow}
\def \ua{\uparrow}
\def \nn{\nonumber}
\def \nnl{\nonumber\\}

\begin{document}

\title{High  Jaynes-Cummings pseudospins eigenstates  in the homogeneous Tavis-Cummings model}
%Homogeneous Tavis-Cummings model as a set of high pseudospin particles interacting with a single bosonic mode
%Rabi overtones
%\title{Analytical solutions in the system of many qubits resonantly coupled through a resonator}

\author{Marcin Dukalski$^a$ and Yaroslav M. Blanter$^{a,b}$}
\affiliation{$^a$Kavli Institute of Nanoscience, Delft University of Technology, Lorentzweg 1, 2628 CJ Delft, The Netherlands; $^b$ Kavli Institute for Theoretical Physics China, CAS, Beijing 100190, China}
\begin{abstract}
We show that a set of $N$ identical qubits coupled  to a single cavity resonator can be decomposed into a set of independent subsystems, which can be thought of as higher pseudospin generalisations of the Jaynes-Cummings model. We derive and analyse the solutions to the equations of motion  and demonstrate unusual beating behaviour resulting from a new form of $\sqrt{n}$-type non-linearity appearing within a pseudospin ladder.
Furthermore, we propose a relative phase shift transformation which allows one to switch/initiate the multi-qubit state in a desired pseudospin configuration, and show how such transformations can be used to undo spontaneous single qubit decay.
We discuss the conditions which justify the validity of the rotating wave approximation in an $N$-qubit system.
%We present an initial Ansatz based method which can be used to calculate the propagator of an arbitrary spin system, and we show how this method can be extended to a system of a single qubit-resonator system with qubit dephasing.
\end{abstract}

% 03.67.Mn 	Entanglement measures, witnesses, and other characterizations (see also
% 03.65.Ud 	Entanglement and quantum nonlocality
 %78.47.jp 	Optical nutation (see also 42.50.Md Optical transient phenomena: quantum beats, photon echo, free-induction decay, dephasings and revivals, optical nutation, and self-induced transparency)
 %78.70.-g 	Interactions of particles and radiation with matter
\maketitle

Superposition and entanglement are among the most striking characteristics that distinguish quantum from classical physics and give way for quantum information processing (QIP) \cite{Nielsen,NazarovBlanterbook}. The building block behind the QIP technologies is a two level system (a qubit), and quantum algorithms rely on coherent manipulation of a single qubit and on controllable coupling between qubits. Recent years brought a tremendous progress in manipulation and coherence control of multiple qubits \cite{Laussy,Arnold}. The control of the individual qubit as well as its read-out is often done by means of coupling it to a resonator (an optical \cite{diamond} or a microwave \cite{MajerCoupling, Fink2} cavity), in the framework of a cavity QED or a circuit QED architecture. In order to use the benefits of the speed-up offered by quantum protocols, one needs to be able to involve a larger number of qubits and place them in a quantum register, to be able to control the coherence as well as the evolution of this multipartite system. Very recently, multiple-qubit registers with three qubits coupled to a single resonator have been fabricated and coherent single-excitation exchange between them was demonstrated \cite{Fink}.

Interaction of a single qubit with a resonator is described by means of the Jaynes-Cummings model \cite{JC}. The behavior of this model is well understood. It predicts such phenomena as the Rabi splitting and Rabi oscillations facilitated by a change in the resonators population, quantum beats, and electromagnetically induced transparency \cite{ScullyZubairy}.

The multipartite analog of the Jaynes-Cummings model, essential for the description of several qubits coupled to the same resonator, is known as the Tavis-Cummings model (TCM) \cite{TCM} and is considerably less understood. The eigenvalues of the Tavis-Cummings Hamiltonian have been found \cite{TCM}, and the full dynamics of the two qubit system has been studied \cite{Man2009}. The inhomogeneous coupling has been addressed \cite{Romero}. The understanding of the model beyond two qubits is very limited \cite{YMO, KochManyQ}, whereas the multitude of excitation schemes available in the model suggests rich and interesting dynamics.

In this Letter, we study TCM for $N$ identical qubits resonantly coupled to a single resonator (bosonic mode). This system is symmetric under any qubit permutations, giving rise to a preferred state description and a system decomposition in terms of higher pseudospin composite particles (highly entangled multi-qubit states). This perception of the problem allows to isolate the dynamics into a finite number of clusters of states and greatly simplifies calculations, reducing the number of dimensions from $2^N$ down to $N+1$. We show that every unique pseudospin particle (a multiplet) is characterized by its own Rabi splitting and that single-qubit Rabi oscillations are replaced by beating-like oscillation patterns. Only at the large number of \emph{excess} resonator excitations these lead to the equidistant splittings observed in NMR experiments on spin multiplets. We demonstrate furthermore that with a single multiple-qubit register one has access to the entire range of individual multiplets by means of conditional phase transformations, and show how this setup can be used to detect and fix single qubit decays (spin flips).

\emph{The system.}
The TCM Hamiltonian reads
\bea
\hat{H}&=&\hat{H}_0+\hat{H}_I\,,\nnl
\hat{H}_0&=&\hbar \om \ahd\ah+\sumlim{i=1}{N}\frac{\hbar\O_i}{2}\s_i^z\,,~~~~
\hat{H}_I=\sumlim{i=1}{N}\hbar g_i \lr{\s_i^+\ah+\s_i^-\ahd}\,.\nn
\eea
Here, the operators $\ahd$, $\ah$ refer to the resonator, and $\s_i$ describe the qubits. Assuming resonant $\om=\O_i \forall i$ and equal coupling, $g_i=g$, we define the interaction picture Hamiltonian
\bea
 \hat{\mathcal{V}}=e^{i \hat{H}_0 t}\hat{H}_I e^{-i \hat{H}_0 t}= g\lr{J_+\ah+J_-\ahd}\,,
\eea
where $J_+=\sumlim{i=1}{N} \s_i^+$,  $J_-=J_+^{\dagger}$, and $\hbar=1$.

The propagator $\hat{U}\lr{t,t'}$, describing the dynamics of the system under the assumption it is closed (very stable qubit and a non-dissipative cavity),  is obtained by multiple integration of the Schr\"{o}dinger equation $i\partial_t\ket{\psi\lr{t}}=\hat{\mathcal{V}}\ket{\psi\lr{t}}$.  For resonant coupling, this yields in the interaction picture $\hat{U}\lr{t,0}=e^{i  \hat{\mathcal{V}} t}=\sumlim{k=0}{\infty}\lr{i  t \hat{\mathcal{V}}}^k/k!$. Direct evaluation of this sum is difficult, as for $N$ qubits the number of dimensions to consider grows exponentially $n_{\rm dim}=2^N$. It would also obscure interesting physics of higher pseudospin particles.

The pseudospin is formed in the following process. A set of $N$ identical qubits has an $N$-fold permutation symmetry: any pair-wise exchange of two qubits does not affect the system dynamics. %, i.e. this set of operations is an isometry group of the Hamiltonian.
%The computational basis spanned by $\ket{00\ldots 00},\ket{00\ldots 01},\ldots,\ket{11\ldots11}$, however, is not the ones optimally describing the system at hand.
The \emph{optimal} (symmetry preserving) basis can be determined from the irreducible representations (irreps) of the permutation group.
%, where larger number of qubits will obviously be isometric under the action of a larger group, in turn characterized by a different irreducible representation.
The simplest example is that of two qubits ($N=2$) spanned by computational basis $\ket{00},\ket{01},\ket{10}$ and $\ket{11}$. These can be combined into a spin singlet $\ket{j=0,m=0}=\lr{\ket{01}-\ket{10}}/\sqrt{2}$ (antisymmetric) and a spin triplet $\ket{j=1,m=1}=\ket{11},\ket{j=1,m=0}=\lr{\ket{01}+\ket{10}}/\sqrt{2},\ket{j=1,m=-1}=\ket{00}$ (symmetric under the exchange of the qubits), where we use the notation adapted from the total angular momentum $j$ and the magnetic $m$ quantum numbers. For $N > 2$ one uses the Young tableaux to construct these states, see the supplementary material. In this formalism, the  states $\ket{j=0,m=0}$, and $\lr{\ket{j=1,m=0~ {\rm or}~ \pm 1}}$ come from the $\scriptsize{\young(1,2)}~\lr{\scriptsize{\young(12)}}$ diagram. % in this description.

%If we express the Hamiltonian in these basis \cite{EntPaperOnTC} we find that the singlet is dark (decouples from the resonator) and the triplet is supported by a fermionic creation and an annihilation operators of a spin-1 particle.
%\bea
%J_{j=1,+}&=& \sqrt{2}\lr{\out{m=1}{m=0}+\out{m=0}{m=-1}}\; ~~~~J_{j=1,-}= J_{j=1,+}^{\dagger}\nnl
%\eea

{\em Three qubits}. The possible diagrams are $\scriptsize\young(123)$, $\scriptsize\young(12,3)$, and $\scriptsize\young(13,2)$, corresponding to a quadruplet and two doublet states, reducing a single problem from $2^3$ dimensions down to three problems in 4, 2, and 2 dimensions. Note that the states obtained from these irreps are not mutually orthogonal, and need further orthogonalisation.

\begin{figure*}
	\centering
		\includegraphics[width=0.95 \textwidth]{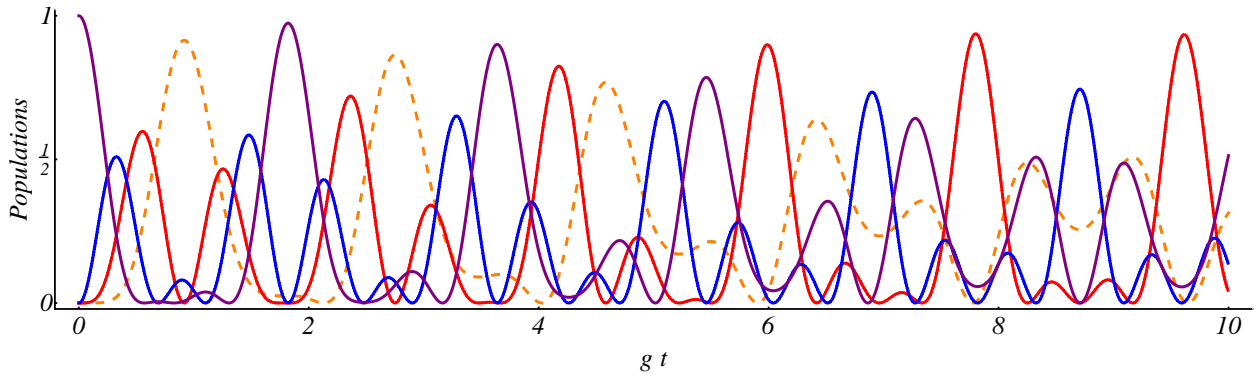}
	\caption{(Color online). Population of the $\ket{m=-\frac{3}{2},4}$ (dashed orange), $\ket{m=-\frac{1}{2},3}$ (dot-dashed red),$\ket{m=\frac{1}{2},2}$ (dotted blue), and $\ket{m=\frac{3}{2},1}$ (continuous purple) states as a function of time (in units inverse coupling strength), given that the state is initiated in the $\ket{m=-\frac{3}{2},n=4}$ state. }
	\label{fig:Allpop}
\end{figure*}

In the supplementary material we present an algorithm for calculating the propagator for arbitrary initial conditions and initial qubits-resonator states correlations. Using this protocol we calculate the three-qubit propagator in the  basis spanned by a quadruplet $\ket{j=\frac{3}{2}}$ and two doublets $\ket{j=\frac{1}{2}}$. The solution to the doublet problem is a subject of standard quantum optics treatments \cite{ScullyZubairy},  and
the propagator for the $j=\frac{3}{2}$ subspace  in the basis  $\ket{j,m}=\left\{\ket{\frac{3}{2},\frac{3}{2}},\ket{\frac{3}{2},\frac{1}{2}},\ket{\frac{3}{2},-\frac{1}{2}},\ket{\frac{3}{2},-\frac{3}{2}}\right\}$ reads
\bea
%\hat{U}\lr{t}=
\lr{\begin{array}{cccc}
 F_{0,1}\lr{\hat{n}} & F_{1,1}\lr{\hat{n}}\ah & F_{2,1}\lr{\hat{n}}\ah^2 & F_{3,1}\lr{\hat{n}}\ah^3 \\
 \ahd F_{1,1}\lr{\hat{n}} & F_{0,2}\lr{\hat{n}} & F_{1,2}\lr{\hat{n}}\ah & F_{2,2}\lr{\hat{n}}\ah^2 \\
 \lr{\ahd}^2 F_{2,1}\lr{\hat{n}} & \ahd F_{1,2}\lr{\hat{n}} & F_{0,3}\lr{\hat{n}} & F_{1,3}\lr{\hat{n}}\ah \\
 \lr{\ahd}^3 F_{3,1}\lr{\hat{n}} & \lr{\ahd}^2 F_{2,2}\lr{\hat{n}} &  \ahd  F_{1,3}\lr{\hat{n}} & F_{0,4}\lr{\hat{n}}
\end{array}}\nn
\eea
where the position (before or after) of the creation and annihilation operators relative to the functions $F_{i,j}$ is very important. Here
\bea
F_{0,1}&=&\sumlim{\pm}{}\lr{\frac{1}{2}\pm\frac{ 2  \hat{n}+7}{ \l_{-,2}-\l_{+,2}}} \cos \lr{\om_ {\pm,2} t} \nn \ ,
\eea
and the rest of these functions as well as the frequencies $\om_ {\pm,2}$ and the coefficients $\l_{\pm,2}$ are given by Eq. (\ref{Fjis3/2sols}) in the supplementary material. The populations of different states are shown in Fig. \ref{fig:Allpop}.

{\em General case}. For $N$ qubits one can always block diagonalise the Hamiltonian, independent of the number of bosonic excitations or the coupling strength.
As a result, we decompose this system with $2^N$ available states into a set of   $^N C_{\left\lfloor\frac{N}{2}\right\rfloor}$, where $\left\lfloor x\right\rfloor$ denotes the floor-function, i.e. one rounding down to the nearest integer.
We label them with $j$ being a (odd half-) consecutive integer valued for an (odd) even number of qubits involved, such that  $\frac{N}{2}\geq j \geq 0 \lr{\frac{1}{2}}$. Moreover, some   $j$ labelled multiplets will be abundant more than others, with the their  abundance ${\rm Ab}_{N,j}$ given by
$$
{\rm Ab}_{N,j}=\frac{\lr{2j+1} N!}{\lr{\frac{N}{2}-j}!\lr{\frac{N}{2}+j+1}!}\,.
$$
The first set of block diagonalising states for two, three and four qubits, are listed in the supplementary material.

Using this form we can predict how many blocks of each type will be present and calculate the within-the-block transition frequencies. The size of the largest $\lr{N+1}\times\lr{N+1}$ block grows linearly, as compared with the exponential growth in the computational basis, as the number of qubits increases.

The $N$-qubit TC Hamiltonian decomposes into a number of high pseudospin Jaynes-Cummings-like terms in the $j^{\rm th}$ representation
\bea
\hat{H}^{TC}_n&=&\bigoplus_{j\in\left\{\frac{\varsigma}{2}\right\}}\lr{\hat{H}^{JC}_j}^{\otimes {\rm Ab}_{n,\varsigma/2}}\,,\nnl
\hat{H}^{JC}_j&=&\delta J_{j,z}+g\lr{J_{j,+}\ah+J_{j,-}\ahd}\,,\nnl
J_{j,+}&=&\sum\limits_{m=-j}^j \tilde{\nu}_{j,m}\out{m+1}{m}\,, ~~~~J_{j,-}= J_{j,+}^{\dagger}\,,\nnl
J_{j,z}&=& \frac{1}{2}\com{J_{j,+}}{J_{j,-}}\,,~~~~\tilde{\nu}_{j,m}=\sqrt{\lr{j-m}\lr{1+j+m}}\,,\nn
\eea
where $\delta$ is the qubit detuning from the resonator and $\varsigma$, $0\leq\varsigma\leq N$, is even (half odd) integer valued for an even (odd) number of  qubits, e.g.
$\hat{H}^{TC}_4=\lr{\hat{H}^{JC}_0}^{\otimes 2} \oplus \lr{\hat{H}^{JC}_1}^{\otimes 3}\oplus \hat{H}^{JC}_2$, or $\hat{H}^{TC}_5= \lr{ \hat{H}^{JC}_{1/2}}^{\otimes 5}\oplus\lr{ \hat{H}^{JC}_{3/2}}^{\otimes 4}\oplus\hat{H}^{JC}_{5/2}$.

\begin{table*}[t]
{\footnotesize
\begin{tabular}{cllll}
$j$-value& $\hat{H}_o$ Degenerate States $\ket{j,n}$&Splitting $[g]$&Large $n$ Splitting $[g]$&$n=0$ Splitting $[g]$\\
$\frac{1}{2}$&$\ket{\frac{1}{2},n},\ket{-\frac{1}{2},n+1}$&$\pm\sqrt{n+1}$&$\approx \pm\sqrt{n}$&$\pm 1$\\
$1$&$\ket{1,n},\ket{0,n+1},\ket{-1,n+2}$&$0,\pm 2\sqrt{n+\frac{3}{2}}$&$\approx 0,\,\pm 2\sqrt{n}$& $0,\pm 2.449$\\
$\frac{3}{2}$&\begin{minipage}{0.3\textwidth}\flushleft$\ket{\frac{3}{2},n},\ket{\frac{1}{2},n+1},$\\$\ket{-\frac{1}{2},n+2},\ket{-\frac{3}{2},n+3}$\end{minipage}&$\pm\sqrt{5\lr{n+2}\pm 4 \sqrt{\lr{n+2}^2+\frac{9}{4}}}$&$\approx \pm\sqrt{n},\,\pm 3\sqrt{n} $ & $\pm 1.207,\pm 4.306$\\
$2$&\begin{minipage}{0.3\textwidth}\flushleft$\ket{2,n},\ket{1,n+1},\ket{0,n+2},$\\$\ket{-1,n+3},\ket{-2,n+4}$\end{minipage}&$0,\pm\sqrt{10\lr{n+\frac{5}{2}}\pm6\sqrt{\lr{n+\frac{5}{2}}^2+2}}$&$\approx 0,\,\pm 2\sqrt{n},\,\pm 4\sqrt{n}$ &$0,\pm 2.787,\pm 6.499$\\
$\frac{5}{2}$&\begin{minipage}{0.3\textwidth}\flushleft$\ket{\frac{5}{2},n},\ket{\frac{3}{2},n+1},\ket{\frac{1}{2},n+2},$\\$\ket{-\frac{1}{2},n+3},\ket{-\frac{3}{2},n+4},\ket{-\frac{5}{2},n+5}$\end{minipage}& No analytical solution&$\approx \pm\sqrt{n},\,\pm 3\sqrt{n},\,\pm 5\sqrt{n} $&$\pm1.364,\pm 4.744,\pm 8.979$\\
%5&\nn
\end{tabular}}
\caption{Summary of Rabi splittings.}\label{rabisplit}
\end{table*}

For simplicity, we choose resonant coupling, $\delta=0$. This reduces the problem of an arbitrary number of \emph{identical} qubits  coupled to a single resonator to the calculation of the propagator for any $j$-representation. Since the Hamiltonian is time independent, the propagator is a $\lr{2j+1} \times \lr{2j+1}$ matrix taking the simple form of $U_{j}\lr{t}=\exp\lr{i t \hat{H}^{JC}_{j}}$. This can be straightforwardly calculated by diagonalisation, exponentiation and inverting the diagonalisation. The eigenvalues of $\hat{H}^{JC}_{j}$ are the new Rabi field splittings, however they no longer correspond to the transition frequencies of the populations between the $\ket{j,m}$ and $\ket{j,m'}$ states, as we saw for propagators for $j=\frac{1}{2}$ or $j=1$ versus $j=\frac{3}{2}$ pseudospin.

\begin{figure}
	\centering
		\includegraphics[width=0.45 \textwidth]{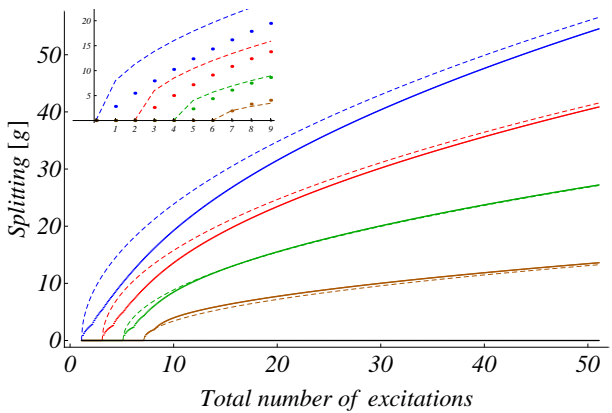}
	\caption{(Color online).  Rabi splittings as a function of the total number of excitations $n$ present in the TCM, for a nonet $\lr{j=4}$ from an 8-qubit register in units of the coupling strength $g$ (dots). We attempt at fitting these trends by a $\lr{8-2k}\sqrt{n-2k}$, for $k=0,1,2,3$ (blue, red,green, and brown respectively). We see that the convergence of the eigenvalues to the fitting trend only works for very high values of $n>30$ and that the convergence is slowest for the highest eigenvalue of $H_{j=4}^{JC}$. Inset zooms in on the few excitations regime. }
	\label{fig:rabisplitfig}
\end{figure}

We write the state of the $j^{\rm th}$ pseudospin interacting with the cavity as $\ket{\psi_j}=\sumlim{i=0}{{\rm Min}\lr{2j,n}} \chi_i \ket{j,m=-j+i}\ket{n-i}$, where the minimum is taken as an upper limit of the sum since for a state ladder composed of $2j+1$ states, there might be fewer than $2j$ photons to climb the ladder. Thus, the dimensionality of the state is restricted by the total initial number of excitations, i.e. if one initialises the system in the state $\ket{j=4,m=-3}\ket{2}$, then the only reachable states are   $\ket{j=4,m=-4}\ket{3}$, $\ket{j=4,m=-3}\ket{2}$, $\ket{j=4,m=-2}\ket{1}$, and $\ket{j=4,m=-1}\ket{0}$. This means that both the pseudospin and the number of available excitations quantitatively and qualitatively affect transitions frequencies between the available states. A list of eigenvalues of the hybridised states are given in {Table \ref{rabisplit}} for $2j\leq n$. In Fig. \ref{fig:rabisplitfig} we show for a nonet ($j=4$ state) how different splittings emerge as the number of available excitations grows.
For $2j\ll n$ (\emph{excess excitations}, i.e. what is left upon climbing from the base to the top of the ladder), all of the spacings of different pseudospins asymptotically converge to  $\pm k g\sqrt{n}$, where $k$ denotes an odd (even) integer, for an odd (even)-half integer valued $j$. In this semi-classical (large $n$) regime the shifted frequencies form a overtones-like spectrum, however in the regime of low $n$, they acquire $\sqrt{n}$-like non-linearities which differ qualitatively for different pseudospin multiplets and also for incompletely used multiplets, i.e. for $2j > n$. These properties are pictured in Fig. \ref{fig:popgrid} in the supplementary material.

%Let us consider a spectrum of excitations which are degenerate on resonance (the resonator energy splitting and the individual qubit
%energy level difference are the same) without the interaction term. For a single qubit we have
%Within this subspace considered, for $g\ll\O,\o$, the interaction term will hybridise the states with new energies. To find these we would need to diagonalise the matrix
%A list of states and the eigenvalues of the hybridised states are given in Table \ref{rabisplit} and the convergence of these eigenvalues for large number of excitations is shown in Figure %\ref{fig:rabisplitfig}.

%Things to be added here
%\begin{itemize}
%\item behaviour - asymptotic, as the number of resonator excess excitations $n \to \infty$ the splittings coalesce and all of the  (half-odd) integer pseudospins will have a similar splitting (which is not observed for direct spin-spin interactions, where the splitting is equidistant to begin with)
%\item emergence of eigenvalues as the number of photons increase, we see that on the graphics grid where the properties change as the number of excitations increases.
%\item show some spectra for many qubits and how they deviate
%\item detection of decay scheme
%\end{itemize}

 \begin{figure*}
	\centering
		\includegraphics[width=0.95 \textwidth]{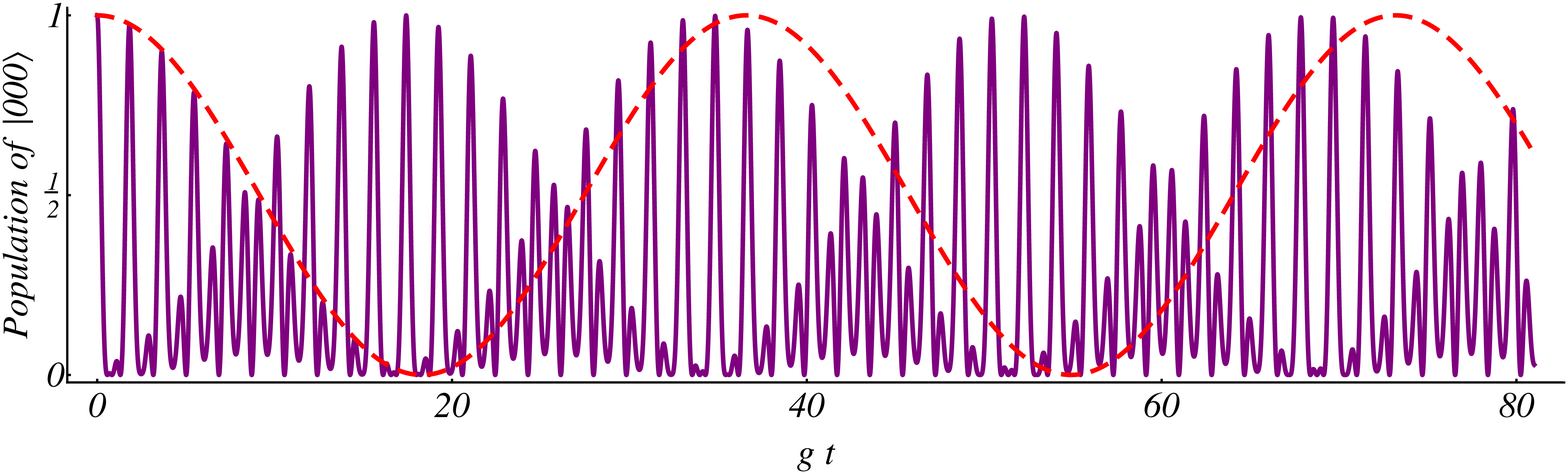}
	\caption{(Color online). Population of the $\ket{000}$ state as a function of time, given that the state is initiated in the $\ket{000,n=4}$ state. We see that the apparent beating does not have a fixed frequency. An attempt to match the beating envelope of this function with a cosine of a single frequency cannot be achieved for an indefinite time interval. }
	\label{fig:pop000}
\end{figure*}

\emph{Switching pseudospins}.
The   permutation symmetry breaks the complete set of states into submanifolds of different pseudospins, which cannot be switched by means of the evolution of the system alone. There exists however a multiply-conditioned phase-shift transformation which allows one to manually change between the submanifolds, by mapping the $j=\frac{N}{2}$ representation state onto a $j=\frac{N}{2}-k$ state. Thus, for $k=1$, in order to reduce $j$ by one, one needs to
initiate the system in the $\ket{000\ldots 0}$ state  ($j=\frac{N}{2}$ manifold) with  one injected bosonic excitation. After $t=\frac{\pi}{2 g \sqrt{N}}$, the state transitions to $\ket{j=\frac{N}{2},m=-\frac{N}{2}+1}$, and by changing the relative phase between the states in the superposition by consecutive $N^{\rm th}$ roots of unity, $\a_N^i$, one forms the $\ket{j=\frac{N}{2}-1,m=-\frac{N}{2}+1}$ state. The application of $J_-$ on that state gives $1/\sqrt{N}\sum_i^N \a_N ^i\ket{000\ldots 0}=0$, meaning that this is the ground state of a new pseudospin ladder.  Transformations for other values of $k$ are more elaborate, with the phase shifts always carried out using $\lr{N-k+1}^{\rm th}$ roots of unity. The details for up to 6 qubits are presented in the supplementary material.

\emph{Effects of decoherence}. For the state-of-the-art realizations, the resonator has coherence times exceeding these of the qubit by over an order of magnitude \cite{Diego}. Therefore we explain here what happens to a three-qubit register if a qubit spontaneously decays.
The system is initiated in the $\ket{000}\otimes\ket{1}$ state and evolves within the symmetric quartet manifold, performing Rabi oscillations with the frequency $g\sqrt{3}$ between the lowest two levels of the state ladder. After a quarter of a period it evolves into the $\ket{j=\frac{3}{2},m=-\frac{1}{2}}\otimes\ket{0}=\frac{1}{\sqrt{3}}\lr{\ket{100}+\ket{010}+\ket{001}}\otimes\ket{0}$ state. At this point the first qubit qubits decays $\r_{before}=\out{j=\frac{3}{2},m=-\frac{1}{2}}{j=\frac{3}{2},m=-\frac{1}{2}}\to \r_{decay}$ resulting in formation of a mixed state
\bea
\r_{decay}&=&\frac{2}{3}\out{\psi}{\psi}\otimes\out{0}{0}+\frac{1}{3}\out{000}{000}\otimes\out{0}{0}\,,\nnl
\ket{\psi}&=&\ket{0}\otimes\lr{\ket{01}+\ket{10}}\nnl
%&=& \frac{1}{\sqrt{6}}\lr{2\ket{j=\frac{3}{2},m=-\frac{1}{2}}-\ket{j=\frac{1}{2}}\otimes \lr{\ket{m=-\frac{1}{2}}+\ket{ m=-\frac{1}{2}}}}\,.\nn
&=& {\small\frac{1}{\sqrt{6}}\lr{2\ket{ \frac{3}{2}, -\frac{1}{2}}- \ket{ \frac{1}{2},-\frac{1}{2}}_1-\ket{ \frac{1}{2}, -\frac{1}{2}}_2}}\,.\nn
\eea
Thus the system splits into two manifolds, one giving rise to the cavity line splitting ($\ket{\frac{3}{2},-\frac{1}{2}}\ket{0}$), and the other one ($\ket{\frac{3}{2},-\frac{3}{2}}\ket{0}$ or $ \lr{\ket{ \frac{1}{2},-\frac{1}{2}}_1+\ket{ \frac{1}{2}, -\frac{1}{2}}_2}\ket{0}/\sqrt{2}$) which does not. Probing the transmittance of the resonator at its bare frequency collapses the state to either the former (no transmittance) with the probability of $\frac{4}{9}$ or the latter (otherwise) set of states ($\frac{5}{9}$). In the latter case, we have two types of states: a dead state $\ket{000}$, with the probability of $P_{\rm dead}=\frac{3}{5}$ and a dormant one (the ground states of the doublets, $P_{\rm dormant}=\frac{2}{5}$). In contrast to the dead state, the dormant one can be woken up. To do that one needs to perform the conditional phase shift operation shifting the states $\left\{j_1=\frac{1}{2},j_2=\frac{1}{2}\right\}\to\left\{j_2=\frac{1}{2},j=\frac{3}{2}\right\}$. Upon a subsequent measurement of the cavity transmittance (at the bare frequency), negative (positive) result collapses the state to the $j=\frac{3}{2}$ $\lr{j=\frac{1}{2}}$ subspace. If the result is positive, one needs to perform one more phase shift operation to recover the initial state.
As a result of this scheme one could undo the decay with the probability of $\frac{2}{3}$, with no need for any direct measurements of the individual qubits. This process can be easily generalised to a greater size qubit register with $\ket{j,m=-j+1}$ states, with the recovery probability as well as the number of potentially needed phase shift increasing with the increasing number of qubits.

For initial states $\ket{000}\otimes\ket{n>1}$, upon a single qubit decay, dead and dormant states will no longer be present, since the ground states of the doublet can interact with the resonator mode, and the newly formed state with $n-1$ total excitations can form a separate quartet manifold of states. This new set of states has different Rabi splittings, relative to a doublet or the quartet with $n$ excitations, allowing this more complex form of $\sqrt{n}$-like non-linearity to be used to detect any decoherence-resulting changes to the system as a whole.

\emph{Strong coupling}. The Jaynes-Cummings type coupling is only a valid description for coupling strengths of the order of $g/\O,\; g/\om\approx 0.1$.  Greater couplings require the use of the counter-rotating terms in the full Rabi model description \cite{Wolf, Braak}. The transition between the $N$ spins state $\ket{000...0}\otimes \ket{1}_{\rm boson}$ and the state with a single excitation in the spin ensemble leads to a $\sqrt{N}$ enhancement of the spin-boson coupling strength \cite{Kurucz}. For a spin-$j$ particle the single boson transition amplitude enhancement $\nu_{j,m}$ depends on the location in the ladder, which ranges from $\sqrt{2j}=\sqrt{N}$, for the states  $\ket{\pm j}$ and $\ket{\pm j\mp 1}$, to $\approx \frac{n}{2}$ for $\ket{m=0}$ and $\ket{m=\pm 1}$ for even, or $\ket{m=\frac{1}{2}}$ and $\ket{m=- \frac{1}{2}}$ for odd $N$. Therefore, using state-of-the-art superconducting qubits with $g/\O\approx 10^{-2}$ \cite{Diego} one can form a spin-10 particle which still operates at weak coupling \cite{Agarwal, NRabi}.

In conclusion, we have shown that a homogeneous Tavis-Cummings model can be simplified to produce higher order pseudospin interacting with a single bosonic mode. Such interaction, for $N>2$ spins gives rise to a beating- like behavior. Moreover, we demonstrated that using a single fixed TC set-up one can emulate an entire range of pseudospins by means of conditional phase change transformations. We gave examples of how using such system, individual qubit decays can be detected and fixed.

Authors would like to thank Chris Drost for useful discussions.
This work was supported by the Netherlands Foundation for Fundamental Research on Matter (FOM).

\newpage

\onecolumngrid

\appendix

\section{Supplementary Material}

\subsection{I. Details on the orthogonal block diagonalising states creation.}
A Young tableaux is a diagram composed of a collection of left-flushed boxes containing numbers which are strongly increasing down each column and across each row, and every consecutive row having a weakly decreasing number of cells (this is commonly referred to as a standard Young tableaux).
%For more details the reader is advised to consult \cite{Fulton}.
By construction every such Young tableaux corresponds to an irreducible representation of a permutation group, such that this particular representation is symmetric under interchange of the elements in the row and anti-symmetric under the interchange of entries between columns (both conditions need to be satisfied simultaneously). Typically, the Young tableaux come very useful in this analysis and reduction of the problem into further sub-problems by means of block-diagonalisation of a Hamiltonian.

  Also here, for a homogeneous Tavis-Cummings model, with a $2^N\times 2^N$ dimensional Hamiltonian, the very convenient multi-qubit states are those given by the irreducible representations of the permutation group. All tableaux that will be used can have at most two rows because that is the greatest number of ways in which we can anti-symmetrise the states.
We will see that the states corresponding to different shapes of the Young tableaux  will always be orthogonal, however states with the same number of qubit excitation   corresponding to the same diagram shape, but different Young tableaux, will no longer be orthogonal. That is why later we will show how a deviation from the Young tableaux formalism further simplifies the problem.

For 2 qubits, there are only two possible Young tableaux, one that corresponds to a triplet state, and another that denotes a dark singlet state.
\bea
\begin{array}{lcrclcr}
\ket{\scriptsize\young(12)\;,m=1}&=&
\ket{11}&~~~~&\ket{\scriptsize\young(12)\;,m=0}&=&\frac{1}{\sqrt{2}}\lr{\ket{01}+\ket{10}}\\
\ket{\scriptsize\young(12)\;,m=-1}&=&\ket{00}&~~~~&
\ket{\scriptsize\young(1,2)\;,m=0}&=&
\frac{1}{\sqrt{2}}\lr{\ket{01}-\ket{10}}
 \end{array}\nn
\eea
where we see that the states that the new set of states forms an orthogonal set of states.

For three qubits there are three possible standard Young tableaux, one perfectly symmetric state $\ket{\scriptsize\young(123)}$, and two states where the third $\ket{\scriptsize\young(12,3)}$ or a second $\ket{\scriptsize\young(13,2)}$ qubit state is antisymmetric , while it is symmetric in the first and second or the first and third respectively.
\bea
\begin{array}{rclcrcl}
\ket{\scriptsize\young(123)\;,m=\frac{3}{2}}&=&\ket{111} &~~~~& \ket{\scriptsize\young(12,3)\;,m=\frac{1}{2}}&=&
 \frac{1}{\sqrt{6}}\lr{2 \ket{110}-\ket{101}-\ket{011}}\\
\ket{\scriptsize\young(123)\;,m=\frac{1}{2}}&=& \frac{1}{\sqrt{3}}\lr{\ket{011}+\ket{101}+\ket{110}} &~~~~&  \ket{\scriptsize\young(12,3)\;,m=-\frac{1}{2}}&=&
 \frac{1}{\sqrt{6}}\lr{2 \ket{001}-\ket{010}-\ket{100}}   \\
\ket{\scriptsize\young(123)\;,m=-\frac{1}{2}}&=& \frac{1}{\sqrt{3}}\lr{\ket{001}+\ket{010}+\ket{100}} &~~~~& \ket{\scriptsize\young(13,2)\;,m=\frac{1}{2}}&=&
\frac{1}{\sqrt{6}}\lr{2 \ket{101} -\ket{110}-\ket{011}}\\
 \ket{\scriptsize\young(123)\;,m=-\frac{3}{2}}&=&\ket{000} &~~~~& \ket{\scriptsize\young(13,2)\;,m=-\frac{1}{2}}&=& \frac{1}{\sqrt{6}}\lr{2 \ket{010} -\ket{001}-\ket{100}}
 \end{array}\nn
\eea
It is easy to see that any state from the $\ket{\scriptsize\young(123)}$ (quadruplet $j=\frac{3}{2}$) manifold is orthogonal to any state in $\ket{\scriptsize\young(1x,y)}$ for $x\neq y \in \left\{2,3\right\}$, however we see that the $\inner{\scriptsize\young(12,3)\;,m_1 }{\scriptsize\young(13,2)\;,m_2 }=-\frac{1}{2}\delta_{m_1m_2}$, indicating that the states are not orthogonal.  This means that the states $\ket{\scriptsize\young(12,3)}$ and $\ket{\scriptsize\young(13,2)}$ would be coupled, and this  is why we propose using alternative orthogonal set of states
\bea
\ket{j=\frac{1}{2}\;,m=\frac{1}{2},1}&=&\frac{1}{\sqrt{3}}\lr{\ket{011}+e^{2\pi i/3}\ket{101}+e^{-2\pi i/3}\ket{110}}\,,\nnl
\ket{j=\frac{1}{2}\;,m=-\frac{1}{2},1}&=&\frac{1}{\sqrt{3}}\lr{\ket{100}+e^{2\pi i/3}\ket{010}+e^{-2\pi i/3}\ket{001}}\,,\nnl
\ket{j=\frac{1}{2}\;,m=\frac{1}{2},2}&=&\frac{1}{\sqrt{3}}\lr{\ket{011}+e^{-2\pi i/3}\ket{101}+e^{2\pi i/3}\ket{110}}\,, \nnl
\ket{j=\frac{1}{2}\;,m=-\frac{1}{2},2}&=&\frac{1}{\sqrt{3}}\lr{\ket{100}+e^{-2\pi i/3}\ket{010}+e^{2\pi i/3}\ket{001}}\,, \nn
\eea
 where the additional (third) label is used to distinguish the different doublet states.

Lastly we just present the states for four qubits
\bea
\begin{comment}
\ket{\scriptsize\young(1234)}&=&\left\{
\begin{array}{c}
 \ket{1111} \\
 \frac{1}{2}\lr{\ket{1110}+\ket{1101}+\ket{1011}+\ket{0111}} \\
 \frac{1}{\sqrt{6}}\lr{\ket{1100}+\ket{1010}+\ket{1001}+\ket{0110}+\ket{0101}+\ket{0011}} \\
 \frac{1}{\sqrt{2}}\lr{\ket{0001}+\ket{0010}+\ket{0100}+\ket{1000}} \\
 \ket{0000}
 \end{array}\right.\nnl
 \end{comment}
\ket{\scriptsize\young(1234)\;,m=2}&=&   \ket{1111} \nnl
\ket{\scriptsize\young(1234)\;,m=1}&=&  \frac{1}{2}\lr{\ket{1110}+\ket{1101}+\ket{1011}+\ket{0111}} \nnl
\ket{\scriptsize\young(1234)\;,m=0}&=&  \frac{1}{\sqrt{6}}\lr{\ket{0011}+\ket{0101}+\ket{0110}+\ket{1001}+ \ket{1010}+\ket{1100}} \nnl
\ket{\scriptsize\young(1234)\;,m=-1}&=&  \frac{1}{2}\lr{\ket{0001}+\ket{0010}+\ket{0100}+\ket{1000}} \nnl
\ket{\scriptsize\young(1234)\;,m=-2}&=&  \ket{0000}\nnl
%%%%%%%%
 \ket{\scriptsize\young(123,4)\;,m=1}&=& \frac{1}{2 \sqrt{3}}\lr{\ket{0111}+\ket{1011}+\ket{1101}-3 \ket{1110}}  \nnl
 \ket{\scriptsize\young(123,4)\;,m=0}&=&\frac{1}{\sqrt{6}}\lr{\ket{0011}+\ket{0101}-\ket{0110}+\ket{1001}-\ket{1010}-\ket{1100}} \nnl
 \ket{\scriptsize\young(123,4)\;,m=-1}&=& \frac{1}{2 \sqrt{3}}\lr{3 \ket{0001}-\ket{0010}-\ket{0100}-\ket{1000}}  \nnl
 %%%%%%%
 \ket{\scriptsize\young(124,3)\;,m=1}&=& \frac{1}{2 \sqrt{3}}\lr{\ket{0111}+\ket{1011}-3 \ket{1101}+\ket{1110}}\nnl
 \ket{\scriptsize\young(124,3)\;,m=0}&=& \frac{1}{\sqrt{6}}\lr{\ket{0011}-\ket{0101}+\ket{0110}-\ket{1001}+\ket{1010}-\ket{1100}}\nnl
 \ket{\scriptsize\young(124,3)\;,m=-1}&=& \frac{1}{2 \sqrt{3}}\lr{\ket{0001}-3 \ket{0010}+\ket{0100}+\ket{1000}}\nnl
%%%%%%%%%
 \ket{\scriptsize\young(134,2)\;,m=1}&=&  \frac{1}{2 \sqrt{3}}\lr{\ket{0111}-3 \ket{1011}+\ket{1101}+\ket{1110}} \nnl
 \ket{\scriptsize\young(134,2)\;,m=0}&=&  \frac{1}{\sqrt{6}} \lr{-\ket{0011}+\ket{0101}+\ket{0110}-\ket{1001}-\ket{1010}+\ket{1100}}\nnl
 \ket{\scriptsize\young(134,2)\;,m=-1}&=& \frac{1}{2 \sqrt{3}}\lr{\ket{0001}+\ket{0010}-3 \ket{0100}+\ket{1000}} \nnl
 %%%%%%%%%%%%%%%%%%%%
 \ket{\scriptsize\young(12,34)\;,m=0}&=&\frac{1}{2\sqrt{3}}\lr{2\ket{0011}+2\ket{1100}-\ket{0101}-\ket{1001}-\ket{1010}-\ket{0110}}\nnl
 \ket{\scriptsize\young(13,24)\;,m=0}&=&\frac{1}{2\sqrt{3}}\lr{2\ket{1010}+2\ket{0101}-\ket{0011}-\ket{0110}-\ket{1001}-\ket{1100}}\nn
\eea
and show the states that corresponding to diagrams $\scriptsize\young(abc,d)$ and  $\scriptsize\young(ab,cd)$ can be orthogonalised to the
triplets and the singlets
\bea
\ket{T_1^{\lr{4}},m=1}&=&\frac{1}{2} (\ket{0 1 1 1}-i \ket{1 0 1 1}-\ket{1 1 0 1}+i \ket{1 1 1 0}) \nnl
\ket{T_1^{\lr{4}},m=0}&=& \frac{1}{2} e^{-3\pi i/4} (i \ket{0 0 1 1}-\ket{0 1 1 0}+\ket{1 0 0 1}-i \ket{1 1 0 0}) \nnl
\ket{T_1^{\lr{4}},m=-1}&=& \frac{1}{2} (-i \ket{0 0 0 1}+\ket{0 0 1 0}+i \ket{0 1 0 0}-\ket{1 0 0 0}) \nnl
\ket{T_2^{\lr{4}},m=1}&=& \frac{1}{2} (\ket{0 1 1 1}-\ket{1 0 1 1}+\ket{1 1 0 1}-\ket{1 1 1 0}) \nnl
\ket{T_2^{\lr{4}},m=0}&=& \frac{1}{\sqrt{2}}\lr{\ket{0 1 0 1}-\ket{1 0 1 0}} \nnl
\ket{T_2^{\lr{4}},m=-1}&=& \frac{1}{2} (\ket{0 0 0 1}-\ket{0 0 1 0}+\ket{0 1 0 0}-\ket{1 0 0 0}) \nnl
\ket{T_3^{\lr{4}},m=1}&=& \frac{1}{2} (\ket{0 1 1 1}+i \ket{1 0 1 1}-\ket{1 1 0 1}-i \ket{1 1 1 0}) \nnl
\ket{T_3^{\lr{4}},m=0}&=& \frac{1}{2} e^{-3\pi i/4} (-\ket{0 0 1 1}+i \ket{0 1 1 0}-i \ket{1 0 0 1}+\ket{1 1 0 0}) \nnl
\ket{T_3^{\lr{4}},m=-1}&=& \frac{1}{2} (i \ket{0 0 0 1}+\ket{0 0 1 0}-i \ket{0 1 0 0}-\ket{1 0 0 0}) \nnl
\ket{S_1^{\lr{4}}}&=& \frac{1}{2}   (\ket{0 0 1 1}-\ket{0 1 1 0}-\ket{1 0 0 1}+\ket{1 1 0 0}) \nnl
\ket{S_2^{\lr{4}}}&=& \frac{1}{2 \sqrt{3}}\lr{\ket{0 0 1 1}-2 \ket{0 1 0 1}+\ket{0 1 1 0}+\ket{1 0 0 1}-2 \ket{1 0 1 0}+\ket{1 1 0 0}}\nn
\eea
but one could also choose the singlets to be
\bea
\ket{\tilde{S}_1^{\lr{4}}}&=&\frac{1}{\sqrt{6}}\lr{\ket{0011}+e^{-\frac{2 i \pi }{3}} \ket{0101}+e^{\frac{2 i \pi }{3}} \ket{0110}+e^{\frac{2 i \pi }{3}} \ket{1001}+e^{-\frac{2 i \pi }{3}} \ket{1010}+\ket{1100}} \nnl
\ket{\tilde{S}_2^{\lr{4}}}&=&\frac{1}{\sqrt{6}}\lr{\ket{0011}+e^{\frac{2 i \pi }{3}} \ket{0101}+e^{-\frac{2 i \pi }{3}} \ket{0110}+e^{-\frac{2 i \pi }{3}} \ket{1001}+e^{\frac{2 i \pi }{3}} \ket{1010}+\ket{1100}} \nn
\eea
We show how to construct these states in general (more than 4 qubits) in the next section.

\begin{table}[t]
\begin{tabular}{r|ccccccccc}
$j$           	&  1  &  2  &  3  &  4  &  5  &  6  &  7  &  8  &  9  \\\hline
 0             	&  0  &  1  &  0  &  2  &  0  &  5  &  0  &  14 &  0  \\
 $\frac{1}{2}$ 	&  1  &  0  &  2  &  0  &  5  &  0  &  14 &  0  &  42 \\
 1   		   	&     &  1  &  0  &  3  &  0  &  9  &  0  &  28 &  0  \\
 $\frac{3}{2}$	&     &     &  1  &  0  &  4  &  0  &  14 &  0  &  48 \\
 2   			&     &     &     &  1  &  0  &  5  &  0  &  20 &  0  \\
 $\frac{5}{2}$	&     &     &     &     &  1  &  0  &  6  &  0  &  27 \\
 3   			&     &     &     &     &     &  1  &  0  &  7  &  0  \\
 $\frac{7}{2}$	&     &     &     &     &     &     &  1  &  0  &  8  \\
 4   			&     &     &     &     &     &     &     &  1  &  0  \\
 $\frac{9}{2}$  &     &     &     &     &     &     &     &     &  1  \\
\end{tabular}
\caption{The number of distinct multiplets depending on the number of a single cavity coupled qubits.}
\end{table}

\subsection{Switching between pseudospins} We now show that one can define a unitary operation that allows one to switch between different types of pseudospin subsystems in the $N$ qubits-resonator system. There is a general transformation which maps the $j=\frac{N}{2}$ representation state onto a $j=\frac{N}{2}-1$ state. Initiating the system in the $\ket{000\ldots 0}$ state, the state will belong to the $j=\frac{N}{2}$ multiplet (it will be the $\ket{j,m=-j}$ state). Provided one excitation is present in the resonator, the state $\ket{j=\frac{N}{2},m=-\frac{N}{2}}$ will transfer to $\ket{j=\frac{N}{2},m=-\frac{N}{2}+1}=1/\sqrt{N}\sum \ket{000\ldots 1}$, where the sum is over all of the $^NC_1=N$ permutations.
We  now define a transformation
\bea
\mathcal{T}:\ket{j=\frac{N}{2},m=-\frac{N}{2}+1}&\to & 1/\sqrt{N}\sum \a_N ^i\ket{000\ldots 1}=\ket{j=\frac{N}{2}-1,m=\frac{N}{2}-1}\nn
\eea
where $\a_N $ is the $N^{\rm th}$ root of unity. An application of $J_-$ on that state gives $1/\sqrt{N}\sum_i^N \a_N ^i\ket{000\ldots 0}=0$, since the sum of all $\a_N $ is zero.  This shows that the newly obtained state is a ground state of another, $j=\frac{N}{2}-1$, multiplet.
 In fact there will be $N-1$ such transformations $\mathcal{T}_k:\ket{j=\frac{N}{2},m=-\frac{N}{2}+1}\to 1/\sqrt{N}\sum \a_N ^{ik}\ket{000\ldots 1}=\ket{j=\frac{N}{2}-1,m=\frac{N}{2}-1}_k$ giving rise to $N-1$ set of ground states of $j=\frac{N}{2}-1$.
Following another injection of excitation one can (for sufficiently large $N$) again perform a transition to a state $\ket{j=\frac{N}{2}-1,m=-\frac{N}{2}+2}$ and attempt a similar transformation. These transformations however become slightly more complicated for $N\geq 5$. Alternatively, one can just perform a two-fold excitation on the $j=\frac{N}{2}$ multiplet and perform a conditional phase gate sequence. For every number of qubits $N$ this can be calculated separately, by imposing a transformation Ansatz
\bea
\mathcal{S}:\ket{j=\frac{N}{2},m=-\frac{N}{2}+2}&&\to  \ket{\psi'}=  \sqrt{\frac{2}{N\lr{N-1}}}\sumlim{i}{} c_i\ket{00\ldots 11}\nn
\eea
such that $\lra{c_i}^2=1$ and $J_-\ket{\psi'}=0$. One then obtains an underconstrained system of $N$ linear equations with $\frac{1}{2}N\lr{N-1}$ unknowns, where one of the solutions is a set of $\lr{N-1}^{\rm th}$ roots of unity.

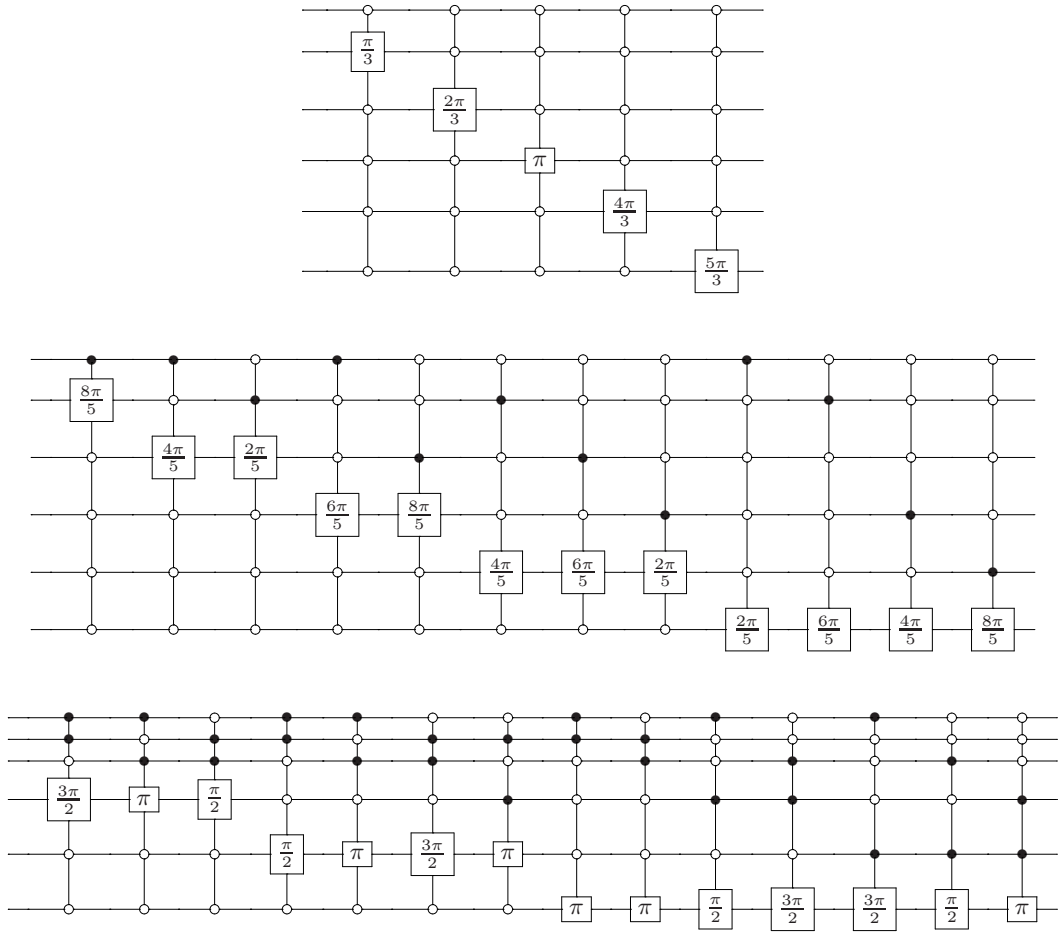
\begin{figure}[b]
\[
\Qcircuit @C=1.0em @R=.7em {
 & \qw & \ctrlo{1}  & \qw & \ctrlo{1}  & \qw & \ctrlo{1}  & \qw & \ctrlo{1}  & \qw & \ctrlo{1} & \qw \\
 & \qw & \gate{\frac{\pi}{3}}   & \qw & \ctrlo{1}  & \qw & \ctrlo{1}  & \qw & \ctrlo{1}  & \qw & \ctrlo{1} & \qw \\
 & \qw & \ctrlo{-1} & \qw & \gate{\frac{2\pi}{3}}   & \qw & \ctrlo{1}  & \qw & \ctrlo{1}  & \qw & \ctrlo{1} & \qw \\
 & \qw & \ctrlo{-1} & \qw & \ctrlo{-1} & \qw & \gate{\pi}   & \qw & \ctrlo{1}  & \qw & \ctrlo{1} & \qw \\
 & \qw & \ctrlo{-1} & \qw & \ctrlo{-1} & \qw & \ctrlo{-1} & \qw & \gate{\frac{4\pi}{3}}   & \qw & \ctrlo{1} & \qw \\
 & \qw & \ctrlo{-1} & \qw & \ctrlo{-1} & \qw & \ctrlo{-1} & \qw & \ctrlo{-1} & \qw & \gate{\frac{5\pi}{3}}  & \qw
}
\]

\[
\Qcircuit @C=0.8em @R=.6em {
 & \qw & \ctrl{1}   &  \qw & \ctrl{1}  & \qw & \ctrlo{1}  & \qw & \ctrl{1}   & \qw &    \ctrlo{1}  & \qw      & \ctrlo{1}
 & \qw & \ctrlo{1}  & \qw & \ctrlo{1}  & \qw & \ctrl{1}   & \qw & \ctrlo{1}  &   \qw & \ctrlo{1}  & \qw & \ctrlo{1}  & \qw \\
 & \qw & \gate{\frac{8\pi}{5}}   & \qw & \ctrlo{1}  & \qw & \ctrl{1}   & \qw & \ctrlo{1}  & \qw  & \ctrlo{1}  & \qw     & \ctrl{1}
 & \qw & \ctrlo{1}  & \qw & \ctrlo{1}  & \qw & \ctrlo{1}  & \qw & \ctrl{1}  &  \qw & \ctrlo{1}  & \qw & \ctrlo{1}  & \qw \\
 & \qw & \ctrlo{-1}  & \qw & \gate{\frac{4\pi}{5}}   & \qw & \gate{\frac{2\pi}{5}}   & \qw & \ctrlo{1}  & \qw   & \ctrl{1}   & \qw   & \ctrlo{1}
 & \qw & \ctrl{1}   & \qw & \ctrlo{1}  & \qw & \ctrlo{1}  & \qw & \ctrlo{1}     & \qw & \ctrlo{1}  & \qw & \ctrlo{1}  & \qw \\
 & \qw & \ctrlo{-1}  & \qw & \ctrlo{-1}  & \qw & \ctrlo{-1}  & \qw & \gate{\frac{6\pi}{5}}   & \qw   & \gate{\frac{8\pi}{5}}   & \qw   & \ctrlo{1}
 & \qw & \ctrlo{1}  & \qw & \ctrl{1}   & \qw & \ctrlo{1}  & \qw & \ctrlo{1}  &   \qw & \ctrl{1}   & \qw & \ctrlo{1}  & \qw \\
 & \qw & \ctrlo{-1}  & \qw & \ctrlo{-1}  & \qw & \ctrlo{-1}  & \qw & \ctrlo{-1}  & \qw   & \ctrlo{-1}  & \qw   & \gate{\frac{4\pi}{5}}
 & \qw & \gate{\frac{6\pi}{5}}   & \qw & \gate{\frac{2\pi}{5}}   & \qw & \ctrlo{1}  & \qw & \ctrlo{1}   & \qw & \ctrlo{1}  & \qw & \ctrl{1}   & \qw \\
 & \qw & \ctrlo{-1}  & \qw & \ctrlo{-1}  & \qw & \ctrlo{-1}  & \qw & \ctrlo{-1}  & \qw   & \ctrlo{-1}  & \qw   & \ctrlo{-1}
 & \qw & \ctrlo{-1}  & \qw & \ctrlo{-1}  & \qw & \gate{\frac{2\pi}{5}}   & \qw & \gate{\frac{6\pi}{5}}      & \qw & \gate{\frac{4\pi}{5}}   & \qw & \gate{\frac{8\pi}{5}}   & \qw
}
\]

\[
\Qcircuit @C=0.8em @R=.5em {
   & \qw & \ctrl{1} & \qw & \ctrl{1} & \qw & \ctrlo{1} & \qw & \ctrl{1} & \qw  & \ctrl{1} & \qw & \ctrlo{1} &   \qw & \ctrlo{1} &     \qw  & \ctrl{1} &   \qw & \ctrlo{1} & \qw & \ctrl{1} & \qw   & \ctrlo{1} & \qw & \ctrl{1} & \qw  & \ctrlo{1} & \qw & \ctrlo{1} & \qw  \\
  & \qw & \ctrl{1} & \qw & \ctrlo{1} & \qw & \ctrl{1} & \qw & \ctrl{1} & \qw  & \ctrlo{1} & \qw & \ctrl{1} & \qw   & \ctrl{1} &   \qw  & \ctrl{1} &   \qw & \ctrl{1} & \qw & \ctrlo{1} & \qw &  \ctrlo{1} & \qw & \ctrlo{1} & \qw & \ctrlo{1} & \qw & \ctrlo{1} & \qw  \\
  & \qw & \ctrlo{1} & \qw & \ctrl{1} & \qw & \ctrl{1} & \qw & \ctrlo{1} & \qw  & \ctrl{1} & \qw & \ctrl{1} &   \qw & \ctrlo{1} &   \qw  & \ctrlo{1} &  \qw & \ctrl{1} & \qw & \ctrlo{1} & \qw &   \ctrl{1} & \qw & \ctrlo{1} & \qw  & \ctrl{1} & \qw & \ctrlo{1} & \qw  \\
   & \qw & \gate{\frac{3\pi}{2}} & \qw & \gate{\pi} & \qw & \gate{\frac{\pi}{2}} & \qw & \ctrlo{1} & \qw  & \ctrlo{1} & \qw & \ctrlo{1} &   \qw & \ctrl{1} &   \qw  & \ctrlo{1} &   \qw & \ctrlo{1} & \qw & \ctrl{1} & \qw   & \ctrl{1} & \qw & \ctrlo{1} & \qw  & \ctrlo{1} & \qw & \ctrl{1} & \qw  \\
   & \qw & \ctrlo{-1} & \qw & \ctrlo{-1} & \qw & \ctrlo{-1} & \qw & \gate{\frac{\pi}{2}} & \qw  & \gate{\pi} & \qw & \gate{\frac{3\pi}{2}} &   \qw & \gate{\pi} &  \qw  & \ctrlo{1} &   \qw & \ctrlo{1} & \qw & \ctrlo{1} & \qw    & \ctrlo{1} & \qw & \ctrl{1} & \qw  & \ctrl{1} & \qw & \ctrl{1} & \qw  \\
  & \qw & \ctrlo{-1} & \qw & \ctrlo{-1} & \qw & \ctrlo{-1} & \qw & \ctrlo{-1} & \qw  & \ctrlo{-1} & \qw & \ctrlo{-1} & \qw   & \ctrlo{-1} &   \qw  & \gate{\pi} &   \qw & \gate{\pi} & \qw & \gate{\frac{\pi}{2}} & \qw    &  \gate{\frac{3\pi}{2}} & \qw & \gate{\frac{3\pi}{2}} & \qw & \gate{\frac{\pi}{2}} & \qw & \gate{\pi} & \qw
}
\]
\caption{A quantum circuit diagram, showing a manipulation of the first second and the third excited septet state to the ground state of the quintet, triplet states or the singlet  state. Here the angles $\theta$ inside the gate boxes denote the phase shift operation.}\label{quantumcircuit2} % $\left(\begin{array}{cc}1&0\\0&e^{i\theta}\end{array}\right)$.}
\end{figure}

This process can be generalised to transformations mapping $\ket{j=\frac{N}{2}} \to \ket{j=\frac{N}{2}-k}$, choosing a  transformed state Ansatz of a superposition of all possible 0s and 1s permutation states with equal amplitudes but different phases. Requiring that the transformed state forms the ground state of a new ladder of states, i.e. $J_- \ket{j=\frac{N}{2}-k}=0$, one obtains a set of $^NC_{k-1}$ equations with $^NC_{k}$ unknowns, leaving a freedom of $^NC_{k-1}-\;^NC_{k}$ parameters. We can always set on of these parameters to 1, since the state can have an arbitrary global phase, and then we will find that among the multitude of possible solutions there is going to be a subset where coefficients $c_i$ are the $\lr{N-k+1}^{\rm th}$ roots of unity.
We present the detailed transformation and the obtained states of a six qubit register in Figure \ref{quantumcircuit2}.

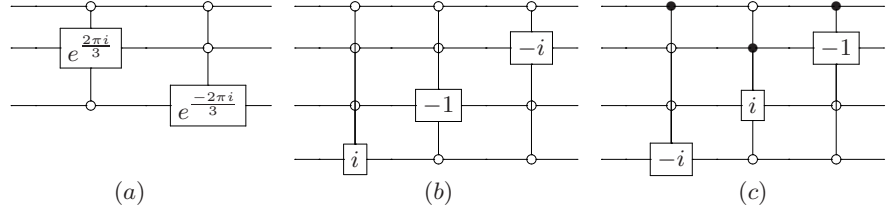
\begin{figure}
\[
\Qcircuit @C=1.0em @R=.7em {
 & \qw &\ctrlo{1} & \qw & \ctrlo{1}  & \qw & ~~~~& \qw &\ctrlo{3} & \qw & \ctrlo{2}  & \qw &\ctrlo{1}  & \qw & ~~~~ & \qw &\ctrl{3}      & \qw & \ctrlo{2}      & \qw &\ctrl{1}       & \qw \\
& \qw &\gate{e^{\frac{2\pi i}{3}}}& \qw  & \ctrlo{1}  & \qw & ~~~~ & \qw &\ctrlo{2} & \qw & \ctrlo{1}  & \qw &\gate{-i}  & \qw & ~~~~ & \qw &\ctrlo{2}     & \qw & \ctrl{1}       & \qw &\gate{-1}      & \qw \\
& \qw &\ctrlo{-1} & \qw & \gate{e^{\frac{-2\pi i}{3}}} & \qw  & ~~~~ & \qw &\ctrlo{1} & \qw & \gate{-1}  & \qw &\ctrlo{-1} & \qw & ~~~~ & \qw &\ctrlo{1}     & \qw & \gate{i}       & \qw &\ctrlo{-1}     & \qw \\
& ~~~~ &~~~~ & ~~~~ & ~~~~ & ~~~~ & ~~~~ & \qw &\gate{i}  & \qw & \ctrlo{-1} & \qw &\ctrlo{-2} & \qw & ~~~~ & \qw &\gate{-i}     & \qw & \ctrlo{-1}     & \qw &\ctrlo{-2}     & \qw \\
 &~~~~ & ~~~~&(a)~~~~ & & & &  & & &(b) & &~~~~~~&&&&& & (c)&
}
\]
\caption{ A quantum circuit diagram, showing a manipulation of (a) the first excited  quartet state to the ground state of the doublet state
(b) the first excited  quintet state to the ground state of the triplet state
(c) the first excited triplet state to the ground state of the singlet state. Here all the gates are the phase change on $\ket{1}$ gates with a phase indicated inside the gate-box.}\label{quantumcircuit}
\end{figure}

\subsection{II. Details of the $N$-qubit propagator calculations}

Here we present two methods of calculating the $j$-particle propagator.  We will first show the method involving the more \emph{purists}' approach, to later present a short-cut.

In the resonant case we can split the propagator into two classes of matrices with an even (odd) parity $\left\{\mathcal{V}_0^a\right\}$ $\lr{\left\{\mathcal{V}_1^b\right\}}$, where $a,b$ are just set indices. These obey $\mathcal{V}_0^a\mathcal{V}_0^b=\mathcal{V}_0^c$,  $\mathcal{V}_1^a\mathcal{V}_1^b=\mathcal{V}_0^c$, and $\mathcal{V}_0^a\mathcal{V}_1^b=\mathcal{V}_1^c$. It is very easy to find the form of these matrices; the even (odd) parity ones have (non) zero entries in the cells where the column and the row digit add up to an odd integer, and vice-versa. Put differently, the even (odd) parity matrices are the even (odd) powers of $ \hat{H}^{JC}_{j}$. In what follows we will show  how to solve an arbitrary qubit number model and then explicitly   solve a three-qubit dynamical model.

In order to find $U_{j}\lr{t}=\exp\lr{i g t \hat{H}^{JC}_{j}/\hbar}=\sumlim{k=0}{\infty}\frac{\lr{i g t}^k}{k!}\lr{\hat{H}^{JC}_{j}}^k$ we have to calculate an arbitrary power of  matrix $\hat{H}^{JC}_{j}$, which is a nontrivial task due to commutation relations of $\ah$ and $\ahd$. This process can however be made simple when using the aforementioned parity decomposition. Using the simplest nontrivial even parity matrix we can then find a recursive relation between the even or odd matrices. Defining $M=\hat{H}^{JC}_{j}$ for shortness, we find that
\bea
M^2&=&\sumlim{p,q=0}{2j+1}\delta_{p,q}\lr{\ah\ahd \tilde{\nu}_{j,p-j}^2+\ahd\ah \tilde{\nu}_{j,p-j-1}^2} + \delta_{p-2,q}\ahd\ahd \tilde{\nu}_{j,q-j}\tilde{\nu}_{j,q-j+1}+\delta_{p,q-2}\ah\ah \tilde{\nu}_{j,p-j}\tilde{\nu}_{j,p-j+1}\nn
\eea
We now form an Ansatz for $M^{2k}$ and  $M^{2k+1}$, $k\in\mathbb{Z}$. In order to avoid the problems with the ordering ambiguity we see that the $\lr{p^{\rm th},q^{\rm th}}$ entry in the propagator must be a process involving a $\lra{p-q}$ photon transfer (absorption if $q>p$ and emission otherwise), moreover this amplitude depends on the number of excitations available. Therefore we can make the following Ansatz
\bea
M^{2k}&=&\sumlim{p,q=1}{2j+1}\delta_{p,q}f_{0,p}^{\lr{2k}}\lr{\hat{n}} +\sumlim{r=1}{2j}\lr{\delta_{p-2r,q}\lr{\ahd}^{2r} f_{2r,q}^{\lr{2k}}\lr{\hat{n}}+\delta_{p,q-2r} f_{2r,p}^{\lr{2k}}\lr{\hat{n}}{\ah}^{2r}}\nnl
M^{2k+1}&=&\sumlim{p,q=1}{2j+1}\sumlim{r=0}{2j}\delta_{p-2r-1,q}\lr{\ahd}^{2r+1} f_{2r+1,q}^{\lr{2k+1}}\lr{\hat{n}} ~~~~~+\delta_{p,q-2r-1} f_{2r+1,p}^{\lr{2k+1}}\lr{\hat{n}}{\ah}^{2r+1}\, ,\nn
\eea
where $f_{\a,p}^{\lr{m}}\lr{\hat{n}}$ is an amplitude, with the labels $\a,p,m$ denoting the number of photons transfered, a function label (as there are many different functions with the same number of photons transferred), and the power of the matrix that these functions appear in.

With this Ansatz we can now use that $M^2M^{2k}=M^{2k}M^2=M^{2(k+1)}$ to find the relationship between these functions. In this way we find that upon equating the first column of the left and the right hand side of this relation we get a recursive relation for the even photon transfer amplitudes in the first column of the propagator
\bea
\vec f^{~2(k+1)}&=&\mathcal{M}\vec f^{~2k}\nnl
\vec f^{~m}&=&\lr{\begin{array}{ccccc}f_{0,1}^{\lr{m}} & f_{2,1}^{\lr{m}}& f_{4,1}^{\lr{m}}& \cdots &f_{2j,1}^{\lr{m}}\end{array}}^T\nnl
\mathcal{M}&=&\sumlim{p,q=0}{j}\delta_{p,q}\lr{(\hat{n}+2p)\tilde{\nu}_{j,2p-j-1}^2+(\hat{n}+1+2p) \tilde{\nu}_{j,p-j}^2} + \delta_{p-1,q} \tilde{\nu}_{j,2q-j}\tilde{\nu}_{j,2q-j+1} +\delta_{p,q-1}\lr{\hat{n}+2q}\lr{\hat{n}+2q-1} \tilde{\nu}_{j,2p-j}\tilde{\nu}_{j,2p-j+1}\nn
\eea

The equation above shows simply a set of linear combination of geometric series, which can be unwound by diagonalising $\mathcal{M}$ and determining it's eigenvalues and using the initial condition $\vec f^{~0}=\sum_{i=1}^{j+1}\delta_{i,1}$. These eigenvalues are then the common ratios and when we sum them up the square-roots of these eigenvalues become the photon-number dependent multi-qubit state Rabi frequencies, however due to a number of them we see that these frequencies form a mutually modulated state transition pattern e.g. $\lr{\hat{U}_{j}\lr{t}}_{\ket{j,j},\ket{j,j}}=\sumlim{k=1}{\left\lfloor j \right\rfloor+1 }A_k \cos{\om_ k t}$. So the diagonalisation of this matrix here is what stands in a way of calculating the eigenvalues and the coefficients in front. This can be done analytically up to four ($j=2$) identical resonantly coupled qubits where there are three frequencies, but one of them is zero (as it is the case every time when an even number of qubits is present). The functions present in the odd parity matrix can be found by identically starting with $M^2M^{2k+1}=M^{2k+1}M^2=M^{2(k+1)+1}$, where the same set of frequencies necessarily needs to be found since otherwise if these non-commensurate frequencies were different in the odd and the even parity sectors, then the total probability would not be conserved which is a necessary condition by construction of the propagator.

In these basis the resonantly coupled interaction picture Hamiltonian reads
\bea
\hat{H}_{j=3/2}=\lr{\begin{array}{llll}
 0 & \sqrt{3} \ah & 0 & 0 \\
 \sqrt{3} \ahd  & 0 & 2 \ah  & 0 \\
 0 & 2 \ahd  & 0 & \sqrt{3} \ah  \\
 0 & 0 & \sqrt{3} \ahd  & 0
\end{array}}\,.
\eea
and as a consequence
\bea
\hat{H}_{j=3/2}^2=M^2=\lr{\begin{array}{llll}
 3 \hat{n}+3 & 0 & 2 \sqrt{3} \ah\ah & 0 \\
 0 & 7 \hat{n}+4 & 0 & 2 \sqrt{3} \ah\ah \\
 2 \sqrt{3} \ahd\ahd & 0 & 7 \hat{n}+3 & 0 \\
 0 & 2 \sqrt{3}  \ahd\ahd & 0 & 3 \hat{n}
\end{array}}
\eea
Let's define an Ansatz for the even parity matrix element
\bea
M^{2k}=\lr{\begin{array}{llll}
 f_{0,1}^{\lr{m}} & 0 & f_{2,1}^{\lr{m}}\ah\ah & 0 \\
 0 & f_{0,2}^{\lr{m}} & 0 & f_{2,2}^{\lr{m}}\ah\ah \\
 \ahd\ahd f_{2,1}^{\lr{m}} & 0 & f_{0,3}^{\lr{m}} & 0 \\
 0 & \ahd\ahd f_{2,2}^{\lr{m}} & 0 & f_{0,4}^{\lr{m}}
\end{array}}
\eea
Then using that $M^{2k}M^{2}=M^{2(k+1)}$ we get that
\bea
f_{0,1}^{\lr{m+1}}&=&3(n+1) f_{0,1}^{\lr{m}}+2\sqrt{3}(n+1)(n+2) f_{2,1}^{\lr{m}}\nnl
f_{2,1}^{\lr{m+1}}&=&2\sqrt{3} f_{0,1}^{\lr{m}}+(17+7n) f_{2,1}^{\lr{m}}\nn
\eea
\emph{I should explain how to get this result}
\bea
f_{0,1}^{\lr{m}}&=& \frac{\lambda_{-}^m (\lambda_{-}-\lambda_{+}-4 \lr{\hat{n}+2}-6)+\lambda_{+}^m (\lambda_{-}-\lambda_{+}+4 \lr{\hat{n}+2}+6)}{2 (\lambda_{-}-\lambda_{+})}\nnl
f_{2,1}^{\lr{m}}&=& \frac{2 \sqrt{3} \left(\lambda_{-}^m-\lambda_{+}^m\right)}{\lambda_{-}-\lambda_{+}}\nnl
\lambda_{\pm}&=&5\lr{\hat{n}+2}\pm\sqrt{16\lr{\hat{n}+2}^2+9}\nn
\eea
There is also an equation which feeds on solutions for $f_{2,1}^{\lr{m}}$ from the above set of equations
\bea
f_{0,3}^{\lr{m+1}}\lr{\hat{n}}=\lr{7n+3}f_{0,3}^{\lr{m}}\lr{\hat{n}}+2\sqrt{3}\ahd\ahd f_{2,1}^{\lr{m}}\lr{\hat{n}}\ah\ah
\eea
where we reinstall the explicit photon number dependence, because now using $g\lr{\hat{n}}\ah=\ah g\lr{\hat{n}-1}$, we can rewrite the equation above
\bea
f_{0,3}^{\lr{m+1}}\lr{\hat{n}}=\lr{7\hat{n}+3}f_{0,3}^{\lr{m}}\lr{\hat{n}}+2\sqrt{3}\lr{\hat{n}-1}\hat{n} f_{2,1}^{\lr{m}}\lr{\hat{n-2}}\nn
\eea
with a solution
\bea
f_{0,3}^{\lr{m}}\lr{\hat{n}}&=&\frac{\varphi_{-}^m (4 \hat{n}+\varphi_{-}-\varphi_{+}+6)-\varphi_{+}^m (4 \hat{n}-\varphi_{-}+\varphi_{+}+6)}{2 (\varphi_{-}-\varphi_{+})}\nnl
\varphi_{\pm}&=&5\hat{n}\pm\sqrt{16\hat{n}^2+9}\nn
\eea

Similarly we have another set of equations
\bea
f_{0,2}^{\lr{m+1}}&=&(7n+4) f_{0,2}^{\lr{m}}+2\sqrt{3}(n+1)(n+2) f_{2,2}^{\lr{m}}\nnl
f_{2,2}^{\lr{m+1}}&=&2\sqrt{3} f_{0,2}^{\lr{m}}+3(n+2) f_{2,2}^{\lr{m}}\nn
\eea
with solutions
\bea
f_{0,2}^{\lr{m}}&=&\frac{\mu_{-}^m (\mu_{-}-\mu_{+}+4 \lr{\hat{n}+1}-6)+\mu_{+}^m (\mu_{-}-\mu_{+}-4  \lr{\hat{n}+1}+6)}{2 (\mu_{-}-\mu_{+})}\nnl
f_{2,2}^{\lr{m}}&=&\frac{2 \sqrt{3} \left(\mu_{-}^m-\mu_{+}^m\right)}{\mu_{-}-\mu_{+}}\nnl
\mu_{\pm}&=&5\lr{\hat{n}+1}\pm\sqrt{16\lr{\hat{n}+1}^2+9}\nn
\eea
and then the finally
\bea
f_{0,4}^{\lr{m+1}}\lr{\hat{n}}&=&3\hat{n}f_{0,4}^{\lr{m}}\lr{\hat{n}}+2\sqrt{3}\ahd\ahd f_{2,2}^{\lr{m}}\lr{\hat{n}}\ah\ah\nnl
f_{0,4}^{\lr{m+1}}\lr{\hat{n}}&=&3\hat{n}f_{0,4}^{\lr{m}}\lr{\hat{n}}+2\sqrt{3}\lr{\hat{n}-1}\hat{n} f_{2,2}^{\lr{m}}\lr{\hat{n-2}}\nn
\eea
with a solution
\bea
f_{0,4}^{\lr{m}}\lr{\hat{n}}&=&\frac{\k_{-}^m (\k_{-}-\k_{+}-4 \lr{\hat{n}-1}+6)+\k_{+}^m (\k_{-}-\k_{+}+4  \lr{\hat{n}-1}-6)}{2 (\k_{-}-\k_{+})}\nnl
\k_{\pm}&=&5\lr{\hat{n}-1}\pm\sqrt{16\lr{\hat{n}-1}^2+9}\nn
\eea

Similarly an the odd subspace is obtained, by first starting with the Ansatz
\bea
M^{2k+1}=\lr{\begin{array}{cccc}
 0 & f_{1,1}^{\lr{m}}\ah & 0 & f_{3,1}^{\lr{m}}\ah^3 \\
 \ahd f_{1,1}^{\lr{m}}  & 0 & f_{1,2}^{\lr{m}}\ah & 0 \\
 0 & \ahd f_{1,2}^{\lr{m}} & 0 &  f_{1,2}^{\lr{m}}\ah\\
 \lr{\ahd}^3 f_{3,1}^{\lr{m}} & 0 & \ahd f_{1,3}^{\lr{m}} & 0
\end{array}}\nnl
\eea
with solutions
\bea
f_{1,1}^{\lr{m}}&=&\frac{\sqrt{3} \left(\lambda _{-}^m (\lambda _{-}-\lambda _{+}+4 p-6)+\lambda _{+}^m (\lambda _{-}-\lambda _{+}-4 p+6)\right)}{2 (7 p-3) (\lambda _{-}-\lambda _{+})}\nnl
f_{3,1}^{\lr{m}}&=&\frac{6 \left(\lambda_{-}^m-\lambda_{+}^m\right)}{(7 p-3) (\lambda_{-}-\lambda_{+})}\frac{6 \left(\lambda _{-}^m-\lambda _{+}^m\right)}{(7 p-3) (\lambda _{-}-\lambda _{+})}\nnl
f_{1,2}^{\lr{m}}&=&\frac{2 \left(\mu _{-}^m-\mu _{+}^m\right)}{\mu _{-}-\mu _{+}}\nnl
f_{1,3}^{\lr{m}}&=&\frac{\varphi _{-}^m (4 q-\varphi _{-}+\varphi _{+}-6)+\varphi _{+}^m (-4 q-\varphi _{-}+\varphi _{+}+6)}{2 \sqrt{3} (q+1) (\varphi _{+}-\varphi _{-})}\nn
\eea
In all of the expressions above, upon summing, every $m^{\rm th}$ power of $\lambda_{\pm},\varphi_{\pm},\mu_{\pm}$ or $\k_{\pm}$ gets replaced by a sine or a cosine of a square-root of this quantity (with some minor changes with the two and three photon entries, since the lowest entries there are missing when summing up).

\bea
 F_{0,1}&=&\frac{(\l_{-,2}+4 \lr{\hat{n}+2}-\l_{+,2}+6) \cos \left(\om_ {+,2} t\right)-(-\l_{-,2}+4 \lr{\hat{n}+2}+\l_{+,2}+6) \cos \left(\om_ {-,2} t\right)}{2 (\l_{-,2}-\l_{+,2})} \nnl
 F_{0,2}&=&\frac{(\l_{-,1}-\l_{+,1}+4 \lr{\hat{n}+1}-6) \cos \left(\om_ {-,1} t\right)+(\l_{-,1}-\l_{+,1}-4 \lr{\hat{n}+1}+6) \cos \left(\om_ {+,1} t\right)}{2 (\l_{-,1}-\l_{+,1})} \nnl
 F_{0,3}&=&\frac{(\l_{-,0}+4 n-\l_{+,0}+6) \cos \left(\om_ {-,0} t\right)-(-\l_{-,0}+4 n+\l_{+,0}+6) \cos \left(\om_ {+,0} t\right)}{2 (\l_{-,0}-\l_{+,0})} \nnl
 F_{0,4}&=&\frac{(-4 \lr{\hat{n}-1}+\l_{-,-1}-\l_{+,-1}+6) \cos \left(\om_ {-,-1} t\right)+(4 \lr{\hat{n}-1}+\l_{-,-1}-\l_{+,-1}-6) \cos \left(\om_ {+,-1} t\right)}{2 (\l_{-,-1}-\l_{+,-1})} \nnl
 F_{2,1}&=&\frac{i \left(\sqrt{\l_{-,2}} (\l_{-,2}-4 \lr{\hat{n}+2}-\l_{+,2}-6) \sin \left(\om_ {-,2} t\right)+\sqrt{\l_{+,2}} (\l_{-,2}+4 \lr{\hat{n}+2}-\l_{+,2}+6) \sin \left(\om_ {+,2} t\right)\right)}{2 \sqrt{3}
   (p-1) (\l_{-,2}-\l_{+,2})} \nnl
 F_{1,2}&=&\frac{2 i \left(\sqrt{\l_{-,1}} \sin \left(\om_ {-,1} t\right)-\sqrt{\l_{+,1}} \sin \left(\om_ {+,1} t\right)\right)}{\l_{-,1}-\l_{+,1}} \nnl
 F_{1,3}&=&\frac{i \left(\sqrt{\l_{-,0}} (\l_{-,0}-4 n-\l_{+,0}+6) \sin \left(\om_ {-,0} t\right)+\sqrt{\l_{+,0}} (\l_{-,0}+4 n-\l_{+,0}-6) \sin \left(\om_ {+,0} t\right)\right)}{2 \sqrt{3}
   (n+1) (\l_{-,0}-\l_{+,0})} \nnl
 F_{2,1}&=&\frac{2 \sqrt{3} \left(\cos \left(\om_ {-,2} t\right)-\cos \left(\om_ {+,2} t\right)\right)}{\l_{-,2}-\l_{+,2}} \nnl
 F_{2,2}&=&\frac{2 \sqrt{3} \left(\cos \left(\om_ {-,1} t\right)-\cos \left(\om_ {+,1} t\right)\right)}{\l_{-,1}-\l_{+,1}} \nnl
 F_{3,1}&=&\frac{i \left(\sqrt{\l_{-,2}} (-\l_{-,2}+10 \lr{\hat{n}+2}+\l_{+,2}) \sin \left(\om_ {-,2} t\right)+\sqrt{\l_{+,2}} (-\l_{-,2}-10 \lr{\hat{n}+2}+\l_{+,2}) \sin \left(\om_ {+,2} t\right)\right)}{3
   \left(p^2-1\right) (\l_{-,2}-\l_{+,2})}\nnl
   \om_ {\pm,x}&=&g\sqrt{\l_{\pm,x}}~~~~\l_{\pm,x}=5\lr{\hat{n}+x}\pm\sqrt{16\lr{\hat{n}+x}^2+9}\label{Fjis3/2sols}
\eea

A short-cut method uses the fact that the system conserves the total number of excitations. Assuming, for now, that we start in a state $\ket{j,m=j,n}$ we see that there is only a finite number of states that will be populated, and every consecutive state down the $j-$ladder will have one more photon. We can then write (using the three qubit example) the Hamiltonian as
\bea
\hat{H}_{j=3/2}=\lr{\begin{array}{llll}
 0 & \sqrt{3} \ah & 0 & 0 \\
 \sqrt{3} \ahd  & 0 & 2 \ah  & 0 \\
 0 & 2 \ahd  & 0 & \sqrt{3} \ah  \\
 0 & 0 & \sqrt{3} \ahd  & 0
\end{array}}\to\hat{H}'_{j=3/2}=\lr{\begin{array}{llll}
 0 & \sqrt{3} \sqrt{n+1} & 0 & 0 \\
 \sqrt{3} \sqrt{n+1}  & 0 & 2 \sqrt{n+2}  & 0 \\
 0 & 2 \sqrt{n+2}  & 0 & \sqrt{3} \sqrt{n+3}  \\
 0 & 0 & \sqrt{3} \sqrt{n+3}  & 0
\end{array}}\,.
\eea
and then the propagator simply takes the form $\exp\lrb{i g t \hat{H}'_{j=3/2}}$, which can be readily calculated using available computer algebra  software. Upon obtaining the result one has to match the expressions found from matrix exponentiation, to the structure of the propagator Ansatz, i.e. identify the entries in the matrix   in the pseudospin space where the creation or an annihilation operator should be placed. Afterwards the propagator needs to be subdivided into regions shown below
\bea
\lr{\begin{array}{llll}
 a & a & a & a \\
 a & b & b & b \\
 a & b & c & c  \\
 a & b & c & d
\end{array}}
\eea
such that the pattern continues for larger matrices introducing regions $e$, $f$, etc.
Subsequently, in the first row and first column (denoted with $a$),  every instance of $\sqrt{n+k}$ (k being an integer) in a numerator should be treated as a creation or annihilation operator, which one needs to place in a  normal ordering convention  i.e. write the annihilation (creation) operators to the right (left) of the number operators. Similarly, in the second region (denoted with $b$), one performs the same process, however after placing the creation and annihilation operators, every remaining instance of variable $n$ needs to be replaced with $n-1$. Then for region $c$ the replacement is $n\to n-2$, for region $d$ it is $n\to n-3$ etc.

\subsection{III.Solutions to the system with dephasing}

The same method as outlined above can be used when studying the solutions to the master equation of a resonator coupled to a  qubit which undergoes dephasing. The equation of such a system in the interaction picture reads
\bea
\dot{\r}=-i  \com{\delta\s_z+g\lr{\s_+\ah+\s_-\ahd}}{\r}+\frac{\phi}{2}\lr{\s_z\r\s_z-\r}\,.
\eea
The perceived difficulty of this problem is two-fold. For one, the system is highly nonlinear, where the number of bosonic operators in a nonlinear fashion affects the transition amplitude of the qubit state (also before we saw $\hat{n}$ present in a square-root as an argument of a sine or a cosine). Secondly, the evolution is no longer unitary and it requires density operator formalism treatment and a use of superoperators acting on the density matrix from both sides simultaneously.

We can address the first problem by taking the density matrix in a general form $\r=\sum\out{i,n}{j,m}$, where $i$ ($n$) and $j$ ($m$) are the qubit (resonator) states, and rewriting it in terms of a product state $\tilde{\r}=\sum\ket{i,n}\ket{j,m}$. As a result of that for any $\hat{A}_1\r\hat{B_2}\to \hat{A}_1\otimes\hat{B_2}^T\tilde{\r}$, where we have adapted a notation where the subscripts denote the space that the operators are acting on
\bea
\ah_1 \ket{i,n}\ket{j,m}&=&\sqrt{n}\ket{i,n-1}\ket{j,m}~\nnl
\ah_2 \ket{i,n}\ket{j,m}&=&\sqrt{m+1}\ket{i,n}\ket{j,m+1}\nnl
\ahd_1 \ket{i,n}\ket{j,m}&=&\sqrt{n+1}\ket{i,n+1}\ket{j,m}\nnl
\ahd_2 \ket{i,n}\ket{j,m}&=&\sqrt{m}\ket{i,n}\ket{j,m-1}\nn
\eea
With this formalism in mind the equation of motion above takes the form
\bea
\dot{\tilde{\r}}=\tilde{H}\tilde{\r}
=\left(
\begin{array}{llll}
 0 & i g \ahd_2  & -i g \ah_1  & 0 \\
 i g \ah_2  & -\phi+i\delta  & 0 & -i g \ah_1  \\
 -i g \ahd_1  & 0 & -\phi-i\delta  & i g \ahd_2  \\
 0 & -i g \ahd_1  & i g \ah_2  & 0
\end{array}
\right)\left(\begin{array}{l} \rho _{11} \\ \rho _{10} \\ \rho _{01} \\ \rho _{00}\end{array}\right)
\eea
with $\r_{ij}$ denoting the qubit space indices of the density matrix, leaving the photonic ones implicit. The general form solution to the equation above reads $\tilde{\r}\lr{t}=\exp\lr{\tilde{H} t}\tilde{\r}\lr{0}$, which again can be found in a closed form provided that we will be able to find $\tilde{H}^k$ for any $k$. For the time being we will analyse the case of $\delta=0$.

 \begin{figure*}
	\centering
		\includegraphics[width=0.95 \textwidth]{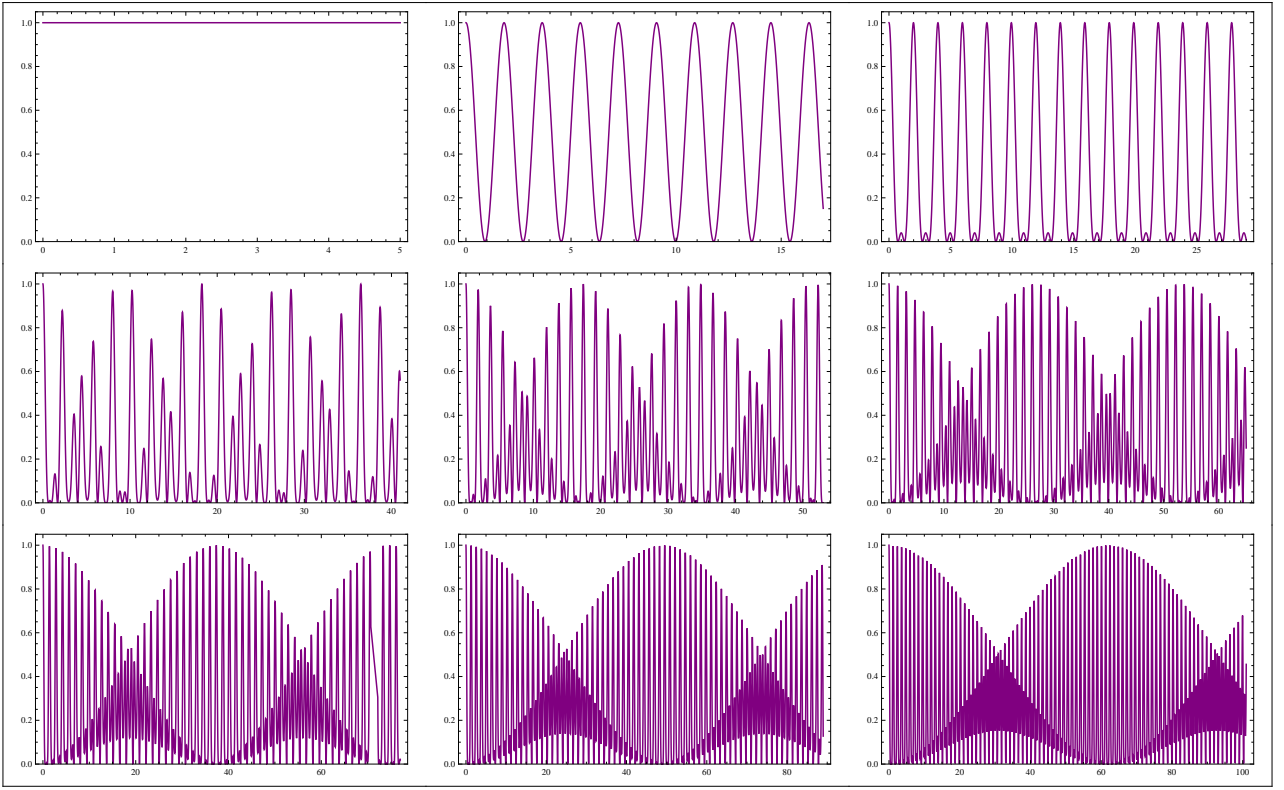}
	\caption{(Color online). Population of the $\ket{000}$ state as a function of time, given that the state is initiated in the $\ket{000,n=k}$ state, where $k=0$ to $k=8$ increasing left to right and down the column. }
	\label{fig:popgrid}
\end{figure*}

By analysing the first few powers of $\tilde{H}$ and using the commutation relations
\bea
\com{\ahd_i}{\ah_j}=\delta_{ij}\lr{-1}^i
\eea
we see that the matrices $\tilde{H}^k$ will always have a particular structure with allowing us to form an Ansatz
\bea
\tilde{H}^k=\left(
\begin{array}{llll}
 f_1\lr{k} & f_5\lr{k} \ahd_2 & f_8\lr{k} \ah_1 &
   f_{10}\lr{k} \ah_1 \ahd_2 \\
 \ah_2 f_5\lr{k} & f_2 & \ah_2 f_6\lr{k} \ah_1 & f_9\lr{k} \ah_1 \\
 \ahd_1 f_8\lr{k} & \ahd_1 f_6\lr{k} \ahd_2 & f_3\lr{k} &
   f_7\lr{k} \ahd_2 \\
 \ahd_1 \ah_2 f_{10}\lr{k} & \ahd_1 f_9\lr{k} &
   \ah_2 f_7\lr{k} & f_4\lr{k}
\end{array}
\right)
\eea
where functions $f_j$ are functions of the number of bosonic excitations $n,m$ in either of the spaces. This allows us to form recursive equations based on $\tilde{H}^{k+1}=\tilde{H}\tilde{H}^k$, which are linearly coupled geometric progressions which can be uncoupled by a simple diagonalisation process.

The solutions $f_j$ to these recursive equations, when summed up to obtain $F_j=\sumlim{k=0}{\infty}\frac{\lr{i f_j}^k}{k!}$ form a product space propagator with the same form as the Ansatz above, but with $f_j\to F_j$, where $F_j$ read

\bea
 F_1&=&\frac{1}{2} e^{-\frac{t \phi }{2}}   (C_{-,1,1}+C_{+,1,1}+S_{-,1,1}+S_{+,1,1}) \nnl
 F_2&=&\frac{1}{2} e^{-\frac{t \phi }{2}}  (C_{-,0,1}+C_{+,0,1}-S_{-,0,1}-S_{+,0,1}) \nnl
 F_3&=& \frac{1}{2} e^{-\frac{t \phi }{2}}   (C_{-,1,0}+C_{+,1,0}-S_{-,1,0}-S_{+,1,0}) \nnl
 F_4&=&\frac{1}{2} e^{-\frac{t \phi }{2}}   (C_{-,0,0}+C_{+,0,0}+S_{-,0,0}+S_{+,0,0}) \nnl
 F_5&=&\frac{ig}{\phi} e^{-\frac{t \phi }{2}}  \left(\eta_{-,1,1}^{1,2} S_{-,1,1}+\eta_{+,1,1}^{1,2}   S_{+,1,1}\right) \nnl
 F_6&=&\frac{e^{-\frac{t \phi }{2}}\lr{C_{-,1,1}-C_{+,1,1}-S_{-,1,1}+S_{+,1,1}}}{{2   \sqrt{(\hat{n}_1+1) (\hat{n}_2+1)}}} \nnl
 F_7&=&\frac{ig}{\phi} e^{-\frac{t \phi }{2}} \lr{\eta_{-,0,1}^{1,2} S_{-,1,0}+\eta_{+,0,1}^{1,2} S_{+,1,0}} \nnl
 F_8&=&-\frac{ig}{\phi} e^{-\frac{t \phi }{2}}  \left(\eta_{-,1,1}^{2,1}   S_{-,1,1}+\eta_{+,1,1}^{2,1} S_{+,1,1}\right) \nnl
 F_9&=& -\frac{i g}{\phi} e^{-\frac{t \phi }{2}}  \left(\eta_{-,0,1}^{2,1} S_{-,0,1}+\eta_{+,0,1}^{2,1} S_{+,0,1}\right) \nnl
 F_{10}&=&\frac{e^{-\frac{t \phi }{2}} \lr{C_{-,1,1}-C_{+,1,1}+S_{-,1,1}-S_{+,1,1}}}{2 \sqrt{(\hat{n}_1+1) (\hat{n}_2+1)}}\nnl
S_{\pm,x,y} &=&\frac{\phi  \sin \lr{\frac{1}{2} g t \lambda_{\pm,x,y}}}{g \lambda_{\pm,x,y}}\nnl
C_{\pm,x,y} &=&  \cos \left(\frac{1}{2} g t \lambda_{\pm,x,y}\right)\nnl
\lambda_{\pm,x,y}&=&\sqrt{4g^2\left(\sqrt{\hat{n}_1+x}\pm\sqrt{\hat{n}_2+y}\right)^2-\phi^2}\nnl
\eta_{\pm,x,y}^{p,q}&=&1\pm\sqrt{\frac{\hat{n}_p+x}{\hat{n}_q+y}}\nn
\eea

In the $t\to\infty$ limit, the propagator above takes a form
\bea
\lim_{t\to\infty} \exp\lr{\tilde{H} t} &=&\frac{1}{2}\lr{\out{\r_{11}}{\r_{11}}+\out{\r_{00}}{\r_{00}}}\nnl
&+&\frac{1}{2}\lr{\frac{\ahd_1\ah_2}{\sqrt{\lr{\hat{n}_1+1}\lr{\hat{n}_2+1}}}\out{\r_{11}}{\r_{00}}+h.c.}\nn
\eea
Assuming that we initialise the cavity-qubit system in a pure state $\r_{\rm init}=\out{\psi}{\psi}$, where $\ket{\psi}=\cos\theta\ket{\ua,n}+e^{i\a}\sin\theta\ket{\da,m}$, then under the evolution of this equation of motion the state will become
\bea
\r_{\rm final}&=&\frac{1}{2}{\rm diag}\left\{c \out{n}{n}+ s \out{m+1}{m+1},\right.\nnl
&&~~~~\left. c\out{n+1}{n+1}+ s\out{m}{m}\right\}\nn
\eea
where $c=\cos^2\theta$ and $s=\sin^2\theta$. This means that regardless of what the initial state of the qubit-cavity system is, the qubit dephasing will lead to loss of any coherence or entanglement within and across the subsystems. Also when we trace out the cavity the qubit is left in the state
$\r_{\rm final}=\frac{1}{2}\lr{\out{\ua}{\ua}+\out{\da}{\da}}$, which is different if the coupling to the resonator was absent $\cos^2\theta\out{\ua}{\ua}+\sin^2\theta\out{\da}{\da}$.

If we treat the detuned case, then based on the same Ansatz and diagonalising the recursive relation, we need to find  the eigenvalues (which are the common ratios of the geometric progression) of the coupled system of functions from the first column of the Ansatz. These eigenvalues later act as the modified Rabi frequencies, and these can be found by means of finding the roots of the characteristic polynomial
\bea
P\lr{x}&=&x^4+2 \phi  x^3+\left(2 n_1 g^2+2 n_2 g^2+4 g^2+\delta ^2+\phi ^2\right) x^2\nnl
  &&+\left(4 \phi  g^2+2 \phi  n_1 g^2+2 \phi  n_2 g^2\right) x
   \phi
\eea
which in the limit of $\delta\to 0$ has the roots $-\phi+i\lambda_{\pm,1,1}$ and its complex conjugate. For non-zero $\delta$, to the best of the authors knowledge, there are no analytical solutions.

\end{document}